\documentclass[aps,showpacs,prd,twocolumn]{revtex4-1}
\usepackage{amsfonts}
\usepackage{amssymb}
\usepackage{amsmath}
\usepackage{color}
\usepackage[usenames,dvipsnames]{xcolor}
\usepackage{float}
\usepackage{accents}
\usepackage{graphicx}
\usepackage{epstopdf}
\usepackage{ulem}
\usepackage[colorlinks=true,pdfstartview=FitV,linkcolor=blue,citecolor=blue,urlcolor=blue,breaklinks=true]{hyperref}

\begin{document}

\title{Nontopological first-order vortices in a gauged $CP(2)$ theory
endowed with the Chern-Simons action}
\author{R. Casana$^{1}$, M. L. Dias$^{1}$ and E. da Hora$^{2}$.}
\affiliation{$^{1}${Departamento de F\'{\i}sica, Universidade Federal do Maranh\~{a}o,}\\
65080-805, S\~{a}o Lu\'{\i}s, Maranh\~{a}o, Brazil.\\
$^{2}$Coordenadoria Interdisciplinar de Ci\^{e}ncia e Tecnologia,\\
Universidade Federal do Maranh\~{a}o, {65080-805}, S\~{a}o Lu\'{\i}s, Maranh\~{a}o, Brazil.}

\begin{abstract}
We study a gauged $CP(2)$ model with the Chern-Simons term, focusing our
attention on those time-independent radially symmetric configurations with
nontopological profile. We proceed the minimization of the effective energy
in order to introduce the corresponding first-order framework, from which we
define a legitimate self-dual scenario. We solve the resulting first-order
equations numerically by means of the finite-difference scheme, from which
we depict the nontopological solutions. We also identify a special kind of
solutions which can be partially described by an analytical treatment.
\end{abstract}

\maketitle

\affiliation{$^{1}${Departamento de F\'{\i}sica, Universidade Federal do Maranh\~{a}o,}\\
65080-805, S\~{a}o Lu\'{\i}s, Maranh\~{a}o, Brazil.\\
$^{2}$Coordenadoria Interdisciplinar de Ci\^{e}ncia e Tecnologia,\\
Universidade Federal do Maranh\~{a}o, {65080-805}, S\~{a}o Lu\'{\i}s, Maranh\~{a}o, Brazil.}

\section{Introduction}

\label{Intro}

In the context of classical field models, vortices are those
time-independent radially symmetric solutions arising from a planar gauged
theory in the presence of a symmetry breaking potential describing the
scalar-matter self-interaction \cite{n5}. However, due to the high
nonlinearity inherent to the symmetry breaking potentials, the corresponding
Euler-Lagrange equations of motion can be quite hard to solve (even
numerically).

On the other hand, under very special circumstances, time-independent
vortices can also be obtained by solving a particular set of coupled
first-order differential equations (instead of the second-order
Euler-Lagrange ones), these equations being usually obtained via the
minimization of the effective energy, the resulting solutions saturating a
well-defined lower bound for the energy itself \cite{n4}.

In this sense, first-order vortices were firstly obtained in the context of
the Maxwell-Higgs electrodynamics in which the corresponding vacuum manifold
exhibits asymmetric states only (the resulting vortices presenting the
typical topological behavior) \cite{n1}. In addition, first-order vortices
were verified to occur also in the Chern-Simons-Higgs theory, with the
vacuum structure presenting now both symmetric and asymmetric states (the
corresponding configurations being topological or nontopological,
respectively) \cite{cshv}.

Furthermore, legitimate vortex solutions satisfying first-order differential
equations were also investigated in connection to the noncanonical gauge
theories \cite{gaht}, the resulting structures being applied in the study of
some interesting cosmological problems \cite{ames}.

In such a context, an interesting issue is the search for the first-order
vortices inherent to a gauged $CP(N-1)$ model, mainly due to the
phenomenological connection between such a theory and the four-dimensional
Yang-Mills-Higgs one \cite{cpn-1}.

In this sense, in a recent work, the time-independent solutions with radial
symmetry arising from a gauged $CP(2)$ model in the presence of the
Maxwell's term were studied, the author focusing his attention on how some
relevant quantities (such as the total energy and the magnetic field) depend
on the parameters defining the model \cite{loginov}. In that work, however,
these configurations were obtained directly from the second-order
Euler-Lagrange equations of motion.

In the sequel, some of us have developed a particular first-order framework
consistent with the very same theoretical scenario described above. Indeed,
we have proceeded the minimization of the resulting energy, from which we
have introduced the corresponding first-order equations and a well-defined
lower bound for the energy itself, the potential supporting self-duality
presenting only asymmetric vacua, which we have used to study
time-independent vortices with a topological profile \cite{casana}.

We have also studied the radially symmetric solitons inherent to a planar $%
CP(2)$ model endowed by the Maxwell's term multiplied by a nontrivial
dielectric function, our main conclusion being that the potential (and the
vacuum manifold it defines) supporting self-duality depends on the dieletric
function itself \cite{lima}. We have then chosen such a function in order to
change the original vacuum manifold into a dot surrounded by a circle (the
centered dot representing a symmetric vacuum), from which we have obtained
nontopological vortices with no electric charge.

Furthermore, we have recently considered a $CP(2)$ theory in the presence of
the Chern-Simons term (instead of the Maxwell's one), via which we have
verified the existence of first-order vortices with a nonvanishing electric
field, the resulting configurations presenting the well-known topological
profile \cite{almeida}.

Now, we go a little bit further on the aforecited investigation by studying
those nontopological vortices satisfying the first-order framework
consistent with the gauged $CP(2)$ model endowed by the Chern-Simons action.

In order to present our results, this work is organized as follows: in the
next Section II, we introduce the overall model and the conventions inherent
to it, focusing our attention on those radially symmetric time-independent
configurations. In the Section III, we split our investigation into two
different branches based on our choices for an additional profile function
which appears in the radially symmetric ansatz. We then proceed the
minimization of the effective energy, from which we introduce the
corresponding first-order framework (i.e. the first-order equations
themselves and a well-defined lower bound for the total energy), the
starting-point being a differential constraint whose solution is the
particular potential supporting self-duality. In the sequel, we use these
expressions in order to define a coherent first-order scenario. We solve the
first-order equations numerically by means of the finite-difference scheme,
from which we depict the solutions to the relevant fields. We also implement
a convenient assumption, from which we get an approximate analytical
description of those numerical solutions, therefore explaining in details
their main properties. In addition, we identify a second type of numerical
solutions that can not be predicted by any analytical treatment. Finally, in
the last Section IV, we present our main conclusions and perspectives
regarding future investigations.

In what follows, we use $\eta ^{\mu \nu }=\left( +--\right) $ as the metric
signature for the flat spacetime, together with the natural units system,
for the sake of convenience.

\section{The model \label{2}}

\label{general}

We begin our letter by reviewing the first-order formalism presented in \cite%
{almeida}, the starting-point being the planar Lagrange density describing
the interaction between the electromagnetic field (introduced via the
Chern-Simons term) and the complex $CP(N-1)$ one, i.e. (here, $\epsilon
^{012}=+1$)%
\begin{equation}
\mathcal{L}=-\frac{\kappa }{4}\epsilon ^{\alpha \mu \nu }A_{\alpha }F_{\mu
\nu }+\left( P_{ab}D_{\mu }\phi _{b}\right) ^{\ast }P_{ac}D^{\mu }\phi
_{c}-V\left( \phi \right) \text{,}  \label{xxm}
\end{equation}%
the $CP(N-1)$ sector itself being constrained to satisfy $\phi _{a}^{\ast
}\phi _{a}=h$. Here,%
\begin{equation}
F_{\mu \nu }=\partial _{\mu }A_{\nu }-\partial _{\nu }A_{\mu }
\end{equation}%
is the electromagnetic field strength tensor and%
\begin{equation}
D_{\mu }\phi _{a}=\partial _{\mu }\phi _{a}-igA_{\mu }Q_{ab}\phi _{b}
\end{equation}%
stands for the usual covariant derivative (in which $Q_{ab}$ is a diagonal
real matrix). Also, $P_{ab}=\delta _{ab}-h^{-1}\phi _{a}\phi _{b}^{\ast }$
is a projection operator.

It is instructive to point out that the theory in (\ref{xxm}) is manifestly
invariant under the global $SU(N)$\ transformation (beyond the usual local $%
U(1)$\ one). In this sense, given that regular solitons are known to occur
during a symmetry breaking phase transition, the first-order scenario we
study in this work is expected to contain a self-interaction potential
depending on only one component of the original $CP(N-1)$\ scalar sector
(therefore giving rise to a spontaneous breaking of the original $SU(N)$\
symmetry), see the discussion in the Ref. \cite{almeida}.

The Euler-Lagrange equation for the Abelian gauge field coming from (\ref%
{xxm}) is%
\begin{equation}
\frac{\kappa }{2}\epsilon ^{\lambda \mu \nu }F_{\mu \nu }=J^{\lambda }\text{,%
}  \label{ggx1}
\end{equation}%
where%
\begin{eqnarray}
J^{\lambda } &=&ig\left[ P_{ac}D^{\lambda }\phi _{c}\left( P_{ab}Q_{bf}\phi
_{f}\right) ^{\ast }\right.  \notag \\
&&\left. ~\ \ \ \ -\left( P_{ab}D^{\lambda }\phi _{b}\right) ^{\ast
}P_{ac}Q_{cb}\phi _{b}\right] \text{,}
\end{eqnarray}%
is the current 4-vector.

It follows from the Eq. (\ref{ggx1}) that the Gauss law for time-independent
configurations reads%
\begin{equation}
\kappa B=\rho \text{,}  \label{xxgl}
\end{equation}%
where $B=F_{21}$ stands for the magnetic field and%
\begin{eqnarray}
\rho &=&g^{2}A^{0}\left[ \left( P_{ab}Q_{ab}\phi _{b}\right) ^{\ast
}P_{ac}Q_{cd}\phi _{d}\right.  \notag \\
&&\left. ~\ \ \ ~\ \ \ -\left( P_{ab}Q_{ab}\phi _{b}\right) \left(
P_{ac}Q_{cd}\phi _{d}\right) ^{\ast }\right] \text{,}
\end{eqnarray}%
represents the stationary charge density. Here, given that $A^{0}=0$ does
not solve (\ref{xxgl}) identically, we conclude that the time-independent
structures arising from (\ref{xxm}) are electrically charged.\ In addition,
in view of the Gauss law (\ref{xxgl}), it is possible to point out that the
total magnetic flux is proportional to the total electric charge, and
vice-versa.

In what follows, we focus our attention on those time-independent radially
symmetric solutions defined by the usual vortex map
\begin{equation}
A_{i}=-\frac{1}{gr}\epsilon ^{ij}n^{j}A(r)\text{,}
\end{equation}%
\begin{equation}
\left(
\begin{array}{c}
\phi _{1} \\
\phi _{2} \\
\phi _{3}%
\end{array}%
\right) =h^{\frac{1}{2}}\left(
\begin{array}{c}
e^{im_{1}\theta }\sin \left( \alpha (r)\right) \cos \left( \beta (r)\right)
\\
e^{im_{2}\theta }\sin \left( \alpha (r)\right) \sin \left( \beta (r)\right)
\\
e^{im_{3}\theta }\cos \left( \alpha (r)\right)%
\end{array}%
\right) \text{,}
\end{equation}%
where $m_{1}$, $m_{2}$ and $m_{3}$ are positive integers defining the
vorticity of the resulting configurations. Also, $\epsilon ^{ij}$ is the
planar Levi-Civita symbol ($\epsilon ^{12}=+1$) and $n^{j}=\left( \cos
\theta ,\sin \theta \right) $ stands for the unit vector. In this case, the
magnetic field can be verified to be given by%
\begin{equation}
B\left( r\right) =-\frac{1}{gr}\frac{dA}{dr}\text{\textbf{,}}  \label{cpo_m}
\end{equation}%
being a function of the radial coordinate $r$\ only.

Here, we point out that regular solutions presenting no divergences are
attained via those profile functions $\alpha (r)$ and $A(r)$ satisfying the
conditions%
\begin{equation}
\alpha (r\rightarrow 0)\rightarrow 0\text{ \ and \ }A(r\rightarrow
0)\rightarrow 0\text{,}  \label{abcx0}
\end{equation}%
which will be used later below. Moreover, given that we are interested in
those first-order solitons with a nontopological profile, the asymptotic
behavior of $\alpha (r)$\ and $A(r)$\ can be supposed to be such as%
\begin{equation}
\alpha \left( r\rightarrow \infty \right) \rightarrow 0\text{ \ and \ }\frac{%
dA}{dr}\left( r\rightarrow \infty \right) \rightarrow 0\text{,}  \label{abcx}
\end{equation}%
with $A_{\infty }\equiv A\left( r\rightarrow \infty \right) $\ finite.

Now, it is important to highlight that, regarding the combination between
the winding numbers $m_{1}$, $m_{2}$ and $m_{3}$ and the charge matrix $%
Q_{ab}$, there are two different choices supporting the existence of
topologically nontrivial configurations (both ones with $m_{3}=0$): (i) $%
m_{1}=-m_{2}=m$ and $Q=\lambda _{3}/2$ (with $\lambda _{3}=$diag$\left(
1,-1,0\right) $), and (ii) $m_{1}=m_{2}=m$ and $Q=\lambda _{8}/2$ ($\sqrt{3}%
\lambda _{8}=$diag$\left( 1,1,-2\right) $). However, it is known that these
two scenarios simply mimic each other, being then phenomenologically
equivalent. Therefore, in this manuscript, we consider only the first
choice, i.e.%
\begin{equation}
Q=\frac{1}{2}\text{diag}\left( 1,-1,0\right) \text{,}
\end{equation}%
with $m_{1}=-m_{2}=m$ and $m_{3}=0$.
\begin{figure}[tbp]
\centering\includegraphics[width=8.5cm]{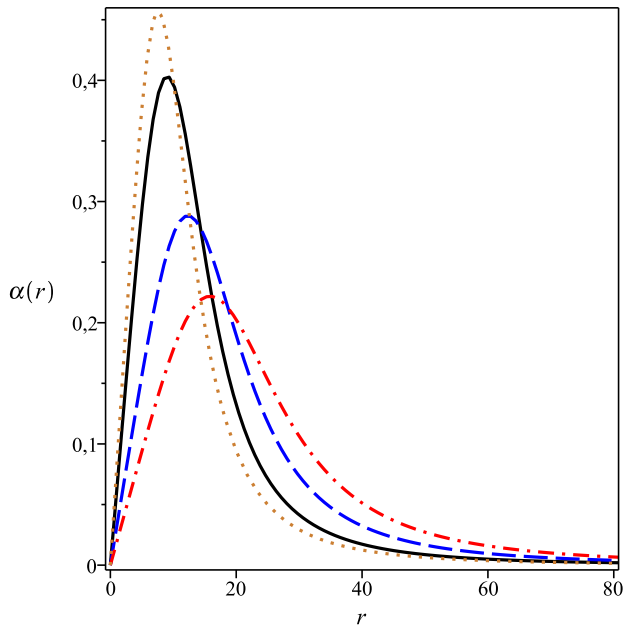}
\par
\vspace{-0.3cm}
\caption{Numerical solutions to $\protect\alpha \left( r\right) $ coming
from the first-order equations (\protect\ref{xbps11}) and (\protect\ref%
{xbps22}) in the presence of the boundary conditions (\protect\ref{abcx0})
and (\protect\ref{abcx}). Here, we have chosen $m=h=g=k=1$ and $r_{0}=10$
(solid black line), $r_{0}=15$ (dashed blue line) and $r_{0}=20$
(dash-dotted red line). We have also plotted the approximate analytical
solution (\protect\ref{asx1}) for $m=h=g=k=1$ and $r_{0}=10$ (dotted orange
line), for comparison. }
\end{figure}

In this case, the radially symmetric Euler-Lagrange equation for the
additional profile function $\beta \left( r\right) $ reads%
\begin{equation}
\frac{d^{2}\beta }{dr^{2}}+\left( \frac{1}{r}+2\cot \alpha \frac{d\alpha }{dr%
}\right) \frac{d\beta }{dr}=H\sin ^{2}\alpha \sin \left( 4\beta \right)
\text{,}
\end{equation}%
where%
\begin{equation}
H(r)=\frac{1}{r^{2}}\left( m-\frac{A}{2}\right) ^{2}-\frac{g^{2}\left(
A_{0}\right) ^{2}}{4}\sin ^{2}\alpha
\end{equation}%
is an auxiliary function, the solutions for $\beta (r)$\ being ($k\in
%TCIMACRO{\U{2124} }%
%BeginExpansion
\mathbb{Z}
%EndExpansion
$)%
\begin{equation}
\beta \left( r\right) =\beta _{1}=\frac{\pi }{4}+\frac{\pi }{2}k\text{ \ \
or \ \ }\beta \left( r\right) =\beta _{2}=\frac{\pi }{2}k\text{,}
\label{betha}
\end{equation}%
this way defining two a priori different scenarios. However, concerning the
first-order configurations, the results for $\beta \left( r\right) =\beta
_{2}$\ can be obtained directly from those for $\beta \left( r\right) =\beta
_{1}$\ via the redefinitions $\alpha \rightarrow 2\alpha $\ and $%
h\rightarrow h/4$, from which it is possible to conclude that there is only
one effective scenario.

We look for the first-order differential equations by proceeding the
minimization of the energy according the Bogomol'nyi prescription, the
starting-point being the energy-momentum tensor itself, i.e.%
\begin{equation}
T_{\lambda \rho }=2\left( P_{ab}D_{\lambda }\phi _{b}\right) ^{\ast
}P_{ac}D_{\rho }\phi _{c}-\eta _{\lambda \rho }\mathcal{L}_{ntop}\text{,}
\label{emt}
\end{equation}%
where%
\begin{equation}
\mathcal{L}_{ntop}=\left( P_{ab}D_{\mu }\phi _{b}\right) ^{\ast
}P_{ac}D^{\mu }\phi _{c}-V\left( \left\vert \phi \right\vert \right)
\end{equation}%
stands for the nontopological sector of the original Lagrange density (\ref%
{xxm}).

The radially symmetric expression for the energy-density coming from (\ref%
{emt}) reads%
\begin{equation}
\varepsilon \left( r\right) =\frac{\kappa ^{2}B^{2}}{g^{2}hW}+h\left[ \left(
\frac{d\alpha }{dr}\right) ^{2}+\frac{W}{r^{2}}\left( \frac{A}{2}-m\right)
^{2}\right] +V\text{,}  \label{edx}
\end{equation}%
where we have used the Gauss law (\ref{xxgl}),%
\begin{equation}
A^{0}=-\frac{2\kappa B}{g^{2}hW}  \label{edxx}
\end{equation}%
in order to rewrite the contribution coming from $A_{0}$\ in terms of the
magnetic field $B$. Here, we have also introduced the auxiliary function%
\begin{equation}
W(\alpha ,\beta )=\left( 1-\sin ^{2}\alpha \cos ^{2}\left( 2\beta \right)
\right) \sin ^{2}\alpha \text{.}
\end{equation}%
It is important to emphasize that, once the function $\beta $\ is assumed to
be a constant (according the values appearing in the Eq. (\ref{betha})), the
potential $V$\ therefore depends on the field $\alpha $\ only, i.e. $%
V=V(\alpha )$.

We also highlight that that the developments we introduce from now on
effectively describe the particular scenario defined by the choices which we
have specified in the previous paragraphs, the solutions for $\beta \left(
r\right) $\ being necessarily one of those in (\ref{betha}).

\section{The solutions \label{2 copy(1)}}

\label{general copy(1)}

\subsection{The BPS formalism for $\protect\beta (r)=\protect\beta _{1}
\displaystyle$}

In view of the discussion right after the Eq. (\ref{betha}), we proceed a
detailed implementation of the first-order BPS formalism for the case
\begin{equation}
\beta \left( r\right) =\beta _{1}=\frac{\pi }{4}+\frac{\pi }{2}k\text{,}
\label{bxx}
\end{equation}%
from which one gets $\cos ^{2}\left( 2\beta _{1}\right) =0$\ and $W\left(
\alpha ,\beta _{1}\right) =\sin ^{2}\alpha $. In this case, the total energy
provided by the expression in (\ref{edx}) then reads%
\begin{eqnarray}
E &=&2\pi \int_{0}^{\infty }\varepsilon \left( r\right) ~rdr  \notag \\
&=&2\pi h\int_{0}^{\infty }\left[ \left( \frac{d\alpha }{dr}\right) ^{2}+%
\frac{\sin ^{2}\alpha }{r^{2}}\left( \frac{A}{2}-m\right) ^{2}\right] rdr
\notag \\
&&+2\pi \int_{0}^{\infty }\left[ \frac{\kappa ^{2}B^{2}}{g^{2}h\sin
^{2}\alpha }+V\right] rdr\text{,}  \label{cdx1}
\end{eqnarray}%
which, after some algebraic manipulations, can be written in the form%
\begin{eqnarray}
E &=&2\pi h\int_{0}^{\infty }\left[ \frac{d\alpha }{dr}\mp \frac{\sin \alpha
}{r}\left( \frac{A}{2}-m\right) \right] ^{2}rdr  \notag \\
&&+2\pi \int_{0}^{\infty }\left( \frac{\kappa B}{g\sqrt{h}\sin \alpha }\mp
\sqrt{V}\right) ^{2}rdr  \notag \\
&&\pm 2\pi \int_{0}^{\infty }\left[ \left( A-2m\right) \frac{h\sin \alpha }{r%
}\frac{d\alpha }{dr}+B\frac{2\kappa \sqrt{V}}{g\sqrt{h}\sin \alpha }\right]
rdr\text{,}
\end{eqnarray}%
or%
\begin{eqnarray}
E &=&2\pi h\int_{0}^{\infty }\left[ \frac{d\alpha }{dr}\mp \frac{\sin \alpha
}{r}\left( \frac{A}{2}-m\right) \right] ^{2}rdr  \notag \\
&&+2\pi \int_{0}^{\infty }\left( \frac{\kappa B}{g\sqrt{h}\sin \alpha }\mp
\sqrt{V}\right) ^{2}rdr  \notag \\
&&\mp 2\pi \int_{0}^{\infty }\left[ (A-2m)h\frac{d\cos \alpha }{dr}\right.
\notag \\
&&\hspace{2cm}\left. +\frac{d(A-2m)}{dr}\frac{2\kappa \sqrt{V}}{g^{2}\sqrt{h}%
\sin \alpha }\right] dr\text{,}  \label{cdx2}
\end{eqnarray}%
where we have used the expression (\ref{cpo_m}) for the magnetic field in
order to write the third row in a convenient form.
\begin{figure}[tbp]
\centering\includegraphics[width=8.5cm]{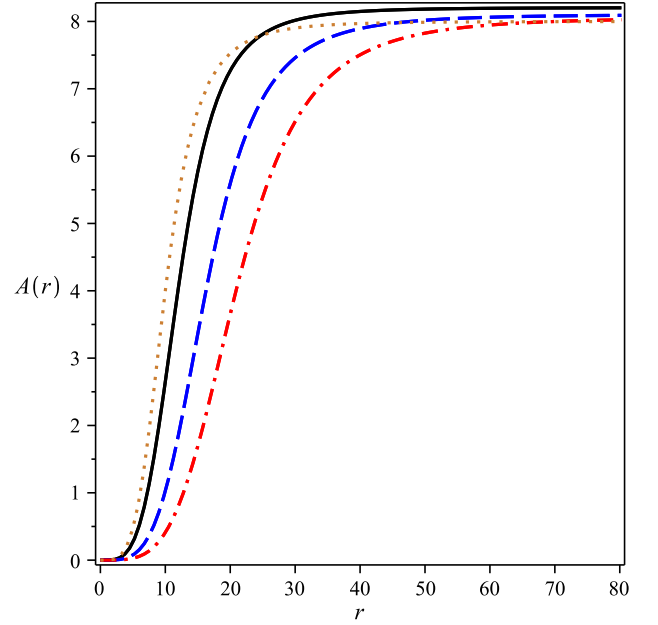}
\par
\vspace{-0.3cm}
\caption{Numerical solutions to $A\left( r\right) $. Conventions as in the
Fig. 1. We have also plotted the approximate analytical solution (\protect
\ref{asx2}). The solutions approach to the value $A_{m}\left( r\rightarrow
\infty \right) =4(m+1)$, the numerical values being $A_{1}\left(
r\rightarrow \infty \right) \approx 8.20526$ for $r_{0}=10$, $A_{1}\left(
r\rightarrow \infty \right) \approx 8.10268$ for $r_{0}=15$ and $A_{1}\left(
r\rightarrow \infty \right) \approx 8.06025$ for $r_{0}=20$.}
\end{figure}

Now, in order to complete the implementation of the first-order BPS
formalism, we need to transform the integrand in the third row in a total
derivative. In this work, we attain such goal by means of the following
relation%
\begin{equation}
\frac{2\kappa }{g^{2}\sqrt{h}}\frac{d}{d\alpha }\left( \frac{\sqrt{V}}{\sin
\alpha }\right) =h\frac{d}{d\alpha }\cos \alpha \text{,}  \label{cx}
\end{equation}%
which also provides the functional form of the self-interacting potential
engendering first-order configurations, i.e.%
\begin{equation}
V\left( \alpha \right) =\frac{g^{4}h^{3}}{16\kappa ^{2}}\sin ^{2}\left(
2\alpha \right) \text{,}  \label{sdx}
\end{equation}%
from which the total energy (\ref{cdx2}) reduces to%
\begin{eqnarray}
E &=&2\pi h\int_{0}^{\infty }\left[ \frac{d\alpha }{dr}\mp \frac{\sin \alpha
}{r}\left( \frac{A}{2}-m\right) \right] ^{2}rdr  \notag \\
&&+2\pi \int_{0}^{\infty }\left( \frac{\kappa B}{g\sqrt{h}\sin \alpha }\mp
\frac{g^{2}h^{3/2}}{4\kappa }\sin \left( 2\alpha \right) \right) ^{2}rdr
\notag \\
&&\mp 2\pi h\int_{0}^{\infty }\frac{d}{dr}\left[ \left( A-2m\right) \cos
\alpha \right] dr\text{.}
\end{eqnarray}

It is instructive to point out that the boundary conditions (\ref{abcx0})
and (\ref{abcx}) allow us to calculate the integral appearing in the third
row explicitly. In this sense, one gets the energy as%
\begin{eqnarray}
E &=&E_{bps}+2\pi h\int_{0}^{\infty }\left[ \frac{d\alpha }{dr}\mp \frac{%
\sin \alpha }{r}\left( \frac{A}{2}-m\right) \right] ^{2}rdr  \notag \\
&&+2\pi \int_{0}^{\infty }\left( \frac{\kappa B}{g\sqrt{h}\sin \alpha }\mp
\frac{g^{2}h^{3/2}}{4\kappa }\sin \left( 2\alpha \right) \right) ^{2}rdr%
\text{,}  \label{EEbps}
\end{eqnarray}%
where%
\begin{equation}
E_{bps}=2\pi \int r\varepsilon _{bps}dr=\mp 2\pi hA_{\infty }  \label{eb}
\end{equation}%
is the lower bound for the energy itself (the Bogomol'nyi bound), the BPS\
energy density $\varepsilon _{bps}$\ standing for%
\begin{equation}
\varepsilon _{bps}=\mp \frac{h}{r}\frac{d}{dr}\left[ \left( A-2m\right) \cos
\alpha \right] \text{.}
\end{equation}

In such a scenario, the Eq. (\ref{EEbps}) tells us that the Bogomol'nyi
bound is saturated when the profile functions satisfy the first-order
differential equations%
\begin{equation}
\frac{d\alpha }{dr}=\pm \frac{\sin \alpha }{r}\left( \frac{A}{2}-m\right)
\text{,}  \label{xbps11}
\end{equation}%
\begin{equation}
B=-\frac{1}{gr}\frac{dA}{dr}=\pm \frac{g^{3}h^{2}}{4\kappa ^{2}}\sin \alpha
\sin \left( 2\alpha \right) \text{,}  \label{xbps22}
\end{equation}%
where the upper (lower) sign holds for negative (positive) values of the
vorticity $m$.
\begin{figure}[tbp]
\centering\includegraphics[width=8.5cm]{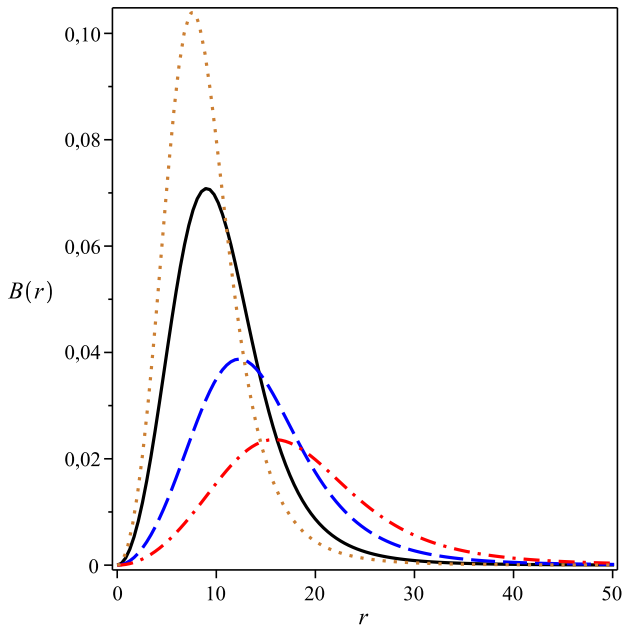} \vspace{-0.3cm}
\caption{Numerical solutions to the magnetic field $B\left( r\right) $.
Conventions as in the Fig. 1. The resulting configurations are rings
centered at the origin, their radii being given by (\protect\ref{raio}). In
particular, $B_{m}\left( r=r_{\max }\right) \propto r_{0}^{-2}$, decreasing
as $r_{0}$ increases.}
\end{figure}

In addition, via the above BPS equations, it is possible to rewrite the
corresponding energy density as%
\begin{equation}
\varepsilon _{bps}=2V(\alpha )+2h\left( \frac{d\alpha }{dr}\right) ^{2}\text{%
,}  \label{EDBPS}
\end{equation}%
with $V(\alpha )$\ being given by the Eq. (\ref{sdx}).

It is interesting to point out that the potential (\ref{sdx}) can be written
in the form%
\begin{equation}
V\left( \left\vert \phi _{3}\right\vert \right) =\frac{g^{4}h}{4k^{2}}%
\left\vert \phi _{3}\right\vert ^{2}\left( h-\left\vert \phi _{3}\right\vert
^{2}\right) \text{,}  \label{potbps}
\end{equation}%
which spontaneously breaks the original $SU(3)$\ symmetry into the $SU(2)$\
one, as expected (see the discussion in the begining of the Section II).

We summarize the overall scenario as follows: once the potential $V(\alpha )$%
\ in (\ref{sdx}) was determined, the profile functions $\alpha (r)$\ and $%
A(r)$\ can be obtained by solving the differential equations (\ref{xbps11})
and (\ref{xbps22}), the resulting radially symmetric configurations
possessing the lowest energy possible, i.e. the Bogomol'nyi bound given by
the Eq. (\ref{eb}).

It is also worthwhile to point out that, concerning the nontopological
configurations we study in this work, the asymptotic contribution appearing
in the energy bound (\ref{eb}) will not be necessarily quantized in terms of
the winding number $m$; this is an essential difference in comparison to the
topological case considered in \cite{almeida}.
\begin{figure}[tbp]
\centering\includegraphics[width=8.5cm]{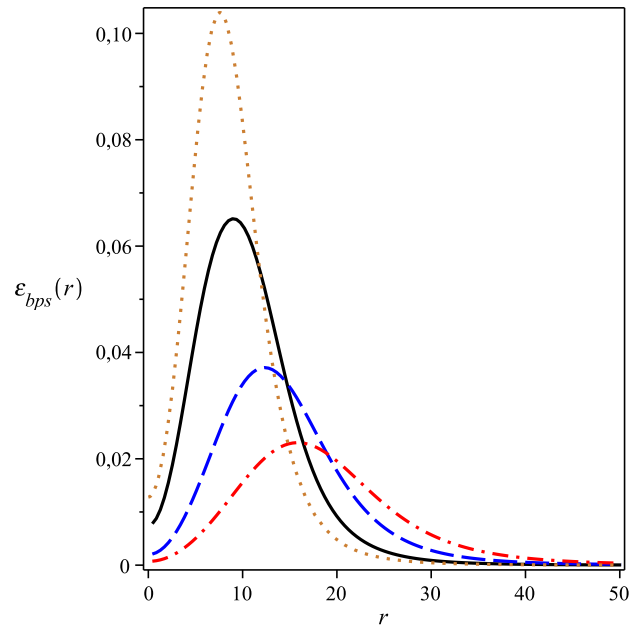}
\par
\vspace{-0.3cm}
\caption{Numerical solutions to the energy density $\protect\varepsilon %
_{bps}(r)$. Conventions as in the Fig. 1. Here, $\protect\varepsilon %
_{bps,m}\left( r=0\right) \propto r_{0}^{-4}$ for $m=1$ and $\protect%
\varepsilon _{bps,m}\left( r=0\right) =0$ for $m>1$.}
\end{figure}

Beyond the BPS energy, other important quantity to be considered is the is
the flux $\Phi _{B}$ of the magnetic field through the planar space,%
\begin{equation}
\Phi _{B}=2\pi \int rB\left( r\right) dr=-\frac{2\pi }{g}A_{\infty }\text{,}
\label{xxmf2}
\end{equation}%
from which one concludes that the energy bound (\ref{eb}) is indeed
proportional to the magnetic flux (\ref{xxmf2}), both quantities being not
necessarily quantized, as expected for nontopological solitons.

\textit{The first-order solutions: the approximate analytical case}. It is
interesting to point out that, due to the conditions $\alpha \left(
r\rightarrow 0\right) \rightarrow 0$\ and $\alpha \left( r\rightarrow \infty
\right) \rightarrow 0$, the first-order equations (\ref{xbps11}) and (\ref%
{xbps22}) can be verified to support approximate analytical solutions. In
order to calculate them, we suppose that $\alpha \left( r\right) \ll 1$ for
all $r$, from which those equations can be approximated, respectively, by%
\begin{equation}
\frac{d\alpha }{dr}=\pm \frac{\alpha }{r}\left( \frac{A}{2}-m\right) \text{,}
\label{abpsx0}
\end{equation}%
\begin{equation}
\frac{1}{r}\frac{dA}{dr}=\mp \frac{g^{4}}{2\kappa ^{2}}h^{2}\alpha ^{2}\text{%
,}  \label{abpsx1}
\end{equation}%
therefore giving rise to Liouville's equation (here, $\lambda
^{2}=g^{4}h^{2}/\kappa ^{2}$)%
\begin{equation}
\frac{d^{2}}{dr^{2}}\ln \alpha ^{2}+\frac{1}{r}\frac{d}{dr}\ln \alpha ^{2}+%
\frac{\lambda ^{2}}{2}\alpha ^{2}=0\text{,}
\end{equation}%
its solution standing for%
\begin{equation}
\alpha \left( r\right) =\frac{4C_{1}}{\lambda r_{0}}\frac{\left( %
\displaystyle\frac{r}{r_{0}}\right) ^{C_{1}-1}}{1+\left( \displaystyle\frac{r%
}{r_{0}}\right) ^{2C_{1}}}\text{,}  \label{xx2}
\end{equation}%
where $r_{0}$ and $C_{1}$ are integration constants. Here, it is worthwhile
to highlight that, in order to fulfill the asymptotic condition $\alpha
\left( r\rightarrow \infty \right) \rightarrow 0$, we must choose $C_{1}>1$.
\begin{figure}[tbp]
\centering\includegraphics[width=8.5cm]{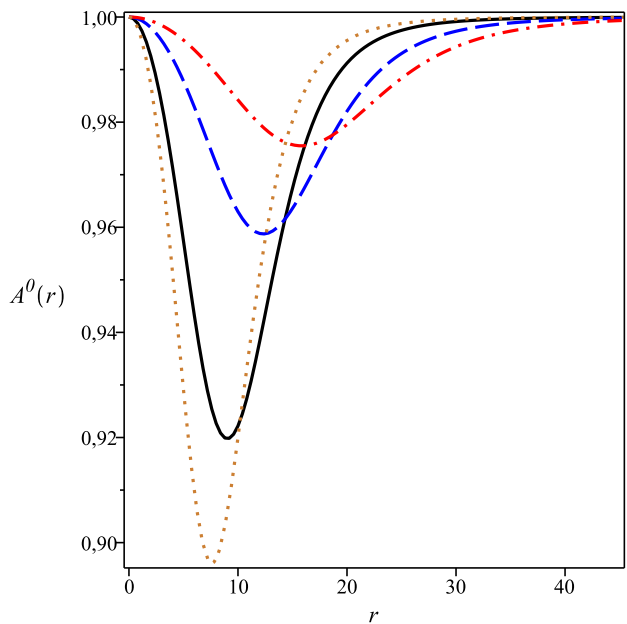}
\par
\vspace{-0.3cm}
\caption{Numerical solutions to the electric potential $A^{0}(r)$.
Conventions as in the Fig. 1. Here, $A_{m}^{0}(r=0)=A_{m}^{0}(r\rightarrow
\infty )=gh/k$, with $A_{m}^{0}(r=r_{\max })$ vanishing for $%
r_{0}\rightarrow \infty $.}
\end{figure}

In addition, given (\ref{abpsx0}) and (\ref{xx2}), one gets that the
solution to $A\left( r\right) $ reads%
\begin{equation}
A\left( r\right) =2\left( m+1-C_{1}\right) +\frac{4C_{1}\left( \displaystyle%
\frac{r}{r_{0}}\right) ^{2C_{1}}}{1+\left( \displaystyle\frac{r}{r_{0}}%
\right) ^{2C_{1}}}\text{,}
\end{equation}%
which satisfies the condition $A\left( r\rightarrow 0\right) \rightarrow 0$
for $C_{1}=m+1$ only.
\begin{figure}[tbp]
\centering\includegraphics[width=8.5cm]{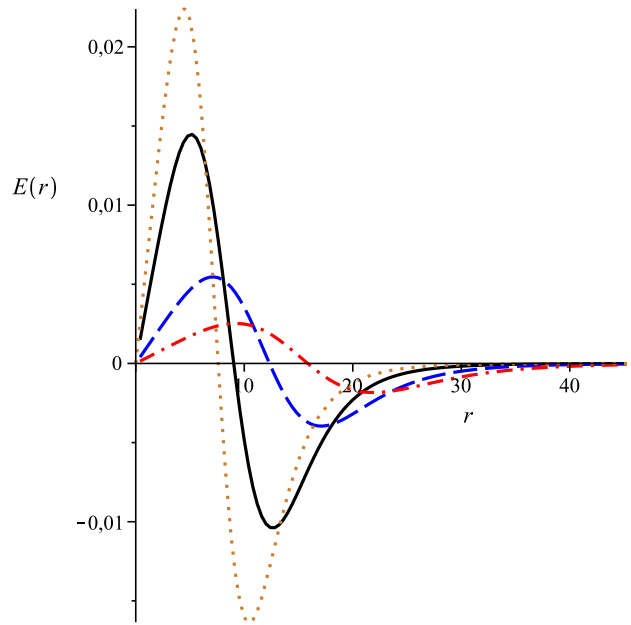}
\par
\vspace{-0.3cm}
\caption{Numerical solutions to the electric field $E(r)$. Conventions as in
the Fig. 1. In this case, $r_{m=1,-}\approx 4.46485$ and $r_{m=1,+}\approx
10.46277$, with $E_{m=1}\left( r=r_{m=1,-}\right) \approx 0.02239$ and $%
E_{m=1}\left( r=r_{m=1,+}\right) \approx -0.01636$. Note the inversion of
the sign dictating the electric interaction.}
\end{figure}

The approximate solutions can then be summarized as%
\begin{equation}
\alpha _{m}\left( r\right) =\frac{4(m+1)}{\lambda r_{0}}\frac{\left( %
\displaystyle\frac{r}{r_{0}}\right) ^{m}}{1+\left( \displaystyle\frac{r}{%
r_{0}}\right) ^{2(m+1)}}\text{,}  \label{asx1}
\end{equation}%
\begin{equation}
A_{m}\left( r\right) =4(m+1)\frac{\left( \displaystyle\frac{r}{r_{0}}\right)
^{2(m+1)}}{1+\left( \displaystyle\frac{r}{r_{0}}\right) ^{2(m+1)}}\text{,}
\label{asx2}
\end{equation}%
the last one giving rise to%
\begin{equation}
A_{m,\infty }\equiv A_{m}\left( r\rightarrow \infty \right) =4(m+1)\text{,}
\end{equation}%
standing for the (approximate) asymptotic condition to be imposed on $%
A\left( r\right) $.

It is interesting to note that the approximate solution to $\alpha
_{m}\left( r\right) $ stands for a well-defined ring, its radius being given
by%
\begin{equation}
r_{\max }=r_{0}\left( \frac{m}{m+2}\right) ^{\frac{1}{2(m+1)}}\text{,}
\label{raio}
\end{equation}%
($r_{\max }$ approaching $r_{0}$ in the limit $m\rightarrow \infty $), from
which one gets%
\begin{equation}
\alpha _{m}\left( r=r_{\max }\right) =\frac{2(m+2)}{\lambda r_{0}}\left(
\frac{m}{m+2}\right) ^{\frac{m}{2(m+1)}}\text{,}  \label{raiox}
\end{equation}%
i.e. the amplitude of the ring, our previous assumption $\alpha \left(
r\right) \ll 1$ holding for%
\begin{equation}
\lambda r_{0}\gg 2(m+2)\left( \frac{m}{m+2}\right) ^{\frac{m}{2(m+1)}}\text{,%
}  \label{raxiox1}
\end{equation}%
i.e., for a fixed $m$, there are only a few values to be chosen for $\lambda
$ and $r_{0}$, and vice-versa.

We have also solved the first-order equations (\ref{xbps11}) and (\ref%
{xbps22}) numerically\ in order to understand the behavior of the profile
fields. In this sense, we have obtained the solutions for $m=h=g=\kappa =1$
and $r_{0}=10$ (solid black line), $r_{0}=15$ (dashed blue line) and $%
r_{0}=20$ (dash-dotted red line), from which we have plotted the resulting
profiles in the figs. 1, 2, 3, 4, 5 and 6 below. We have also depicted the
approximate solutions for $m=h=g=\kappa =1$ and $r_{0}=10$ (dotted orange
line), for comparison.
\begin{figure}[tbp]
\centering\includegraphics[width=8.5cm]{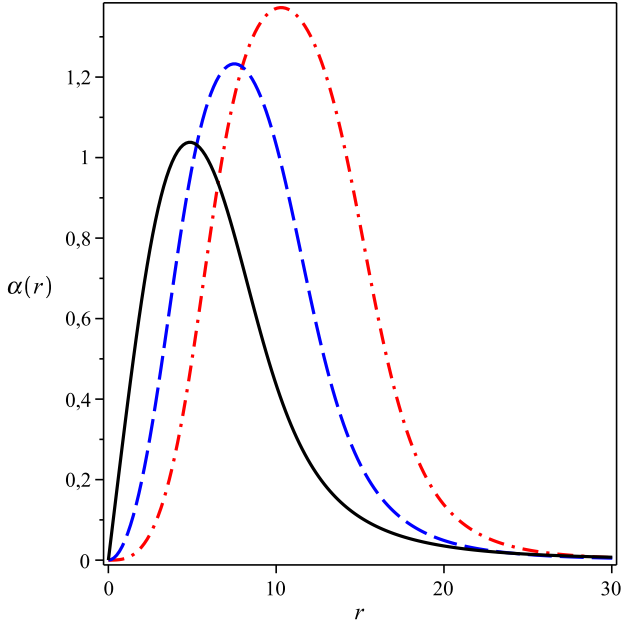}
\par
\vspace{-0.3cm}
\caption{Numerical solutions to $\protect\alpha \left( r\right) $. Here, we
have used $h=g=k=r_{0}=1$ and $m=1$ (solid black line), $m=2$ (dashed blue
line) and $m=3$ (dash-dotted red line), the resulting scenario being not
predictable by any analytical treatment.}
\end{figure}

The solutions to the profile function $\alpha \left( r\right) $ appear in
the Fig. 1. These profiles are well-defined rings centered at the origin,
their radii and amplitudes being given, respectively, by (\ref{raio}) and (%
\ref{raiox}), the first (second) one increasing (decreasing) as $r_{0}$
itself increases.

The Figure 2 shows the numerical results to the profile function $A\left(
r\right) $. Here, it is interesting to note the way the solutions try to
reach the approximate value $A_{m}\left( r\rightarrow \infty \right) =4(m+1)$%
, the true numerical values reading $A_{1}\left( r\rightarrow \infty \right)
\approx 8.20526$ for $r_{0}=10$, $A_{1}\left( r\rightarrow \infty \right)
\approx 8.10268$ for $r_{0}=15$ and $A_{1}\left( r\rightarrow \infty \right)
\approx 8.06025$ for $r_{0}=20$, the overall solutions being monotonic, as
expected.

In the Figure 3, we depict the profiles to the magnetic field $B\left(
r\right) $, the resulting structures also standing for defined rings
centered at $r=0$ (here, both $B\left( r=0\right) $ and $B\left(
r\rightarrow \infty \right) $ vanish). In particular, the approximate
analytical solution to $B_{m}\left( r\right) $ arising from (\ref{asx1}) and
(\ref{asx2}) reads%
\begin{equation}
B_{m}\left( r\right) =\pm \frac{g^{3}h^{2}}{2\kappa ^{2}}\alpha _{m}^{2}%
\text{,}
\end{equation}%
via which one concludes that the radii of the corresponding rings are also
given by (\ref{raio}), the amplitudes being%
\begin{equation}
B_{m}\left( r=r_{\max }\right) =\pm \frac{2g^{3}h^{2}(m+2)^{2}}{\left(
\lambda r_{0}\right) ^{2}\kappa ^{2}}\left( \frac{m}{m+2}\right) ^{\frac{m}{%
m+1}}\text{,}
\end{equation}%
which decrease as $r_{0}$ itself increases.
\begin{figure}[tbp]
\centering\includegraphics[width=8.5cm]{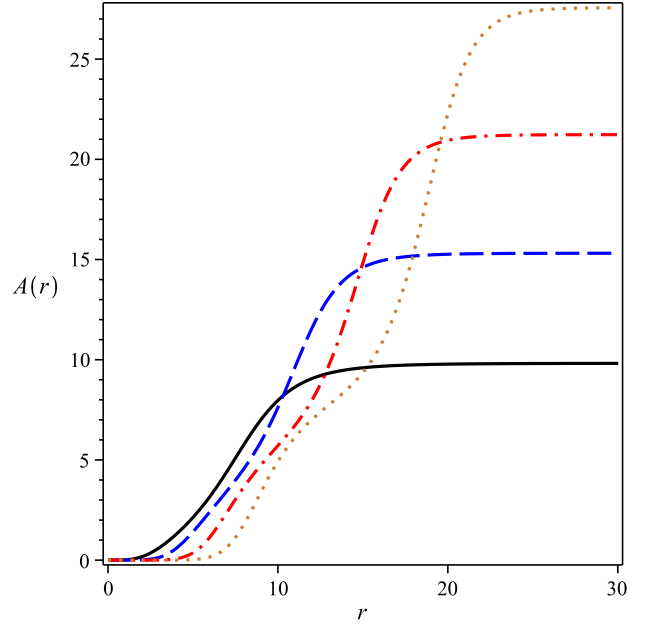}
\par
\vspace{-0.3cm}
\caption{Numerical solutions to $A\left( r\right) $. Conventions as in the
Fig. 7, the dotted orange line representing the solution for $m=4$, the
resulting gauge profile presenting an internal structure.}
\end{figure}

The numerical solutions to the energy density $\varepsilon _{bps}(r)$ are
plotted in the Fig. 4. In this case, it is worthwhile to point out that all
the solutions fulfill the finite-energy requirement, i.e. $\varepsilon
_{bps}\left( r\rightarrow \infty \right) \rightarrow 0$, the approximate
expression for the energy distribution according (\ref{EDBPS}) is
\begin{equation}
\varepsilon _{bps,m}\left( r\right) =\frac{g^{4}h^{3}}{2\kappa ^{2}}\alpha
_{m}^{2}+2h\left( \frac{d\alpha _{m}}{dr}\right) ^{2}\text{,}
\end{equation}%
from which we get that the radii inherent to the energy-rings are also
defined by the expression in (\ref{raio}). Moreover, we calculate%
\begin{equation}
\varepsilon _{bps,m}\left( r=r_{\max }\right) =\frac{2h\left( m+2\right) ^{2}%
}{r_{0}^{2}}\left( \frac{m}{m+2}\right) ^{\frac{m}{m+1}}
\end{equation}%
and%
\begin{equation}
\varepsilon _{bps,m}\left( r=0\right) =\left\{
\begin{array}{c}
\displaystyle{\frac{128m^{2}\kappa ^{2}}{g^{4}hr_{0}^{4}}}\text{, if }m=1 \\
\\
0\text{, if }m>1%
\end{array}%
\right. \text{,}
\end{equation}%
with $\varepsilon _{bps,1}\left( r=0\right) $ decreasing as $r_{0}$
increases, see the numerical solutions.

We plot the numerical results to the electric potential $A^{0}(r)$ in the
Figure 5, the approximate solution standing for%
\begin{equation}
A_{m}^{0}(r)=\pm \frac{gh}{\kappa }\left( 1-\frac{1}{2}\alpha
_{m}^{2}\right) \text{,}  \label{ep}
\end{equation}%
the resulting profile satisfying $A_{m}^{0}(r=0)=A_{m}^{0}(r\rightarrow
\infty )=\pm gh/\kappa $, these boundary conditions do not depending on $m$.
Moreover, given (\ref{ep}), one concludes that the corresponding radius is
also given by (\ref{raiox}), via which we calculate%
\begin{equation}
A_{m}^{0}(r=r_{\max })=\pm \frac{gh}{\kappa }\left( 1-\frac{2(m+2)^{2}}{%
\left( \lambda r_{0}\right) ^{2}}\left( \frac{m}{m+2}\right) ^{\frac{m}{m+1}%
}\right) \text{,}
\end{equation}%
which vanishes in the limit $r_{0}\rightarrow \infty $. In particular, for $%
m=h=g=\kappa =1$ and $r_{0}=10$, one gets that $A_{m=1}^{0}(r=r_{\max
})\approx 0.89608$, see the Figure 5.
\begin{figure}[tbp]
\centering\includegraphics[width=8.5cm]{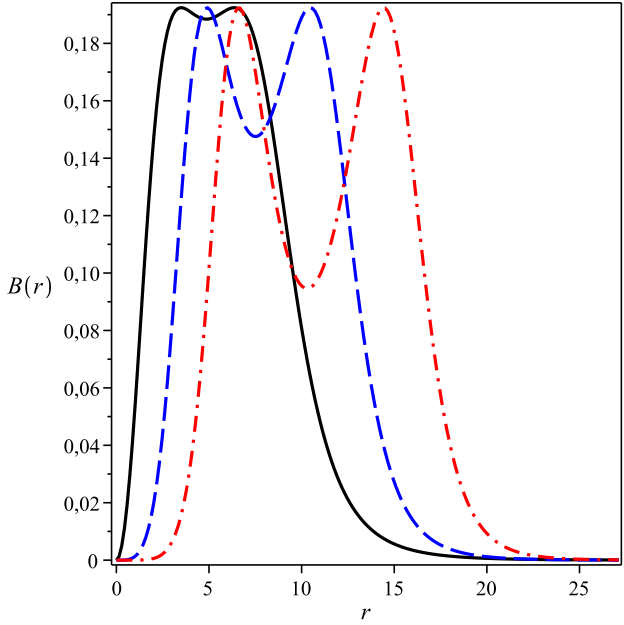}
\par
\vspace{-0.3cm}
\caption{Numerical solutions to $B\left( r\right) $. Conventions as in the
Fig. 7. The solution is a double ring centered at the origin, the magnetic
field vanishing at the boundaries.}
\end{figure}

The numerical solutions to the electric field $E(r)=-dA^{0}/dr$ appear in
the Figure 6, the approximate one reading%
\begin{equation}
E_{m}(r)=\frac{gh}{\kappa }\frac{\alpha _{m}^{2}}{r}\left( \frac{A_{m}}{2}%
-m\right) \text{,}  \label{ef}
\end{equation}%
with $A_{m}(r)$ itself given by (\ref{asx2}). In this case, one gets that%
\begin{equation}
\frac{dE_{m}}{dr}=-\frac{d^{2}A_{m}^{0}}{dr^{2}}=\pm \frac{gh}{\kappa }\left[
\left( \frac{d\alpha _{m}}{dr}\right) ^{2}+\alpha _{m}\frac{d^{2}\alpha _{m}%
}{dr^{2}}\right]
\end{equation}%
vanishes for%
\begin{equation}
\left( \frac{d\alpha _{m}}{dr}\right) ^{2}=-\alpha _{m}\frac{d^{2}\alpha _{m}%
}{dr^{2}}\text{,}
\end{equation}%
whose solutions are%
\begin{equation}
r_{m,\mp }=r_{0}R_{m,\mp }^{\frac{1}{2\left( m+1\right) }}\text{,}
\end{equation}%
in which%
\begin{equation}
R_{m,\mp }=\frac{a_{m}\mp b_{m}}{c_{m}}\text{,}
\end{equation}%
the positive coefficients%
\begin{equation}
a_{m}=4m^{2}+8m+1\text{,}
\end{equation}%
\begin{equation}
b_{m}=\sqrt{12m^{4}+48m^{3}+61m^{2}+26m+1}\text{,}
\end{equation}%
and%
\begin{equation}
c_{m}=2m^{2}+9m+10
\end{equation}%
depending on the vorticity $m$ explicitly.
\begin{figure}[tbp]
\centering\includegraphics[width=8.5cm]{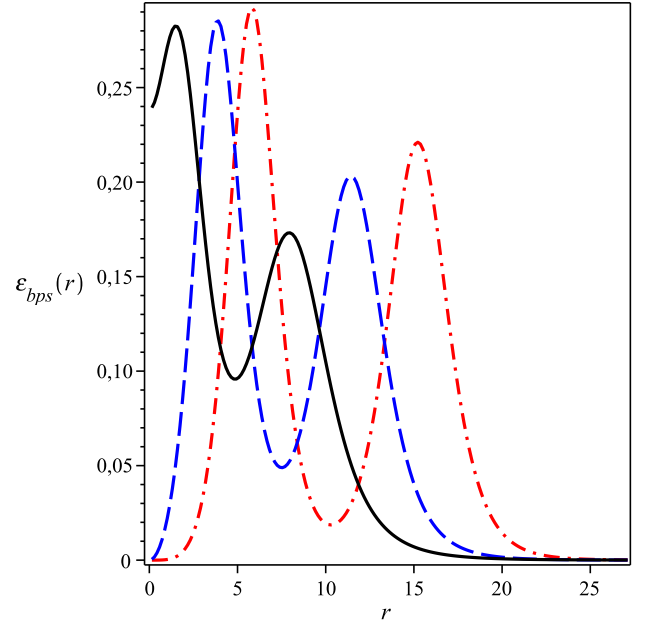}
\par
\vspace{-0.3cm}
\caption{Numerical solutions to $\protect\varepsilon _{bps}\left( r\right) $%
. Conventions as in the Fig. 7, the energy distribution vanishing at $r=0$
for $m\neq 1$ only.}
\end{figure}

In the Figure 6, $r_{m,-}$ and $r_{m,+}$ are the points in which the
approximate solution (\ref{ef}) for the electric field reaches its extreme
values, i.e.%
\begin{equation}
E_{m}\left( r=r_{m,\mp }\right) =\frac{16(m+1)^{2}gh}{\lambda ^{2}\kappa
r_{0}}\Sigma _{m,\mp }
\end{equation}%
for $m>0$, and%
\begin{equation}
E_{m}\left( r=r_{m,\mp }\right) =-\frac{16(m+1)^{2}gh}{\lambda ^{2}\kappa
r_{0}}\Sigma _{m,\mp }
\end{equation}%
for $m<0$, where%
\begin{equation}
\Sigma _{m,\mp }=\frac{R_{\mp }^{\frac{2m-1}{2\left( m+1\right) }}}{\left(
1+R_{\mp }\right) ^{3}}\left( m-\left( m+2\right) R_{\mp }\right) \text{,}
\end{equation}%
with both $E_{m}\left( r=0\right) $ and $E_{m}\left( r\rightarrow \infty
\right) $ vanishing. In particular, again for $m=h=g=\kappa =1$ and $%
r_{0}=10 $, we get that $r_{m=1,-}\approx 4.46485$ and $r_{m=1,+}\approx
10.46277$, with $E_{m=1}\left( r=r_{m=1,-}\right) \approx 0.02239$ and $%
E_{m=1}\left( r=r_{m=1,+}\right) \approx -0.01636$. Here, it is interesting
to note the inversion in the sign dictating the electric interaction.
\begin{figure}[tbp]
\centering\includegraphics[width=8.5cm]{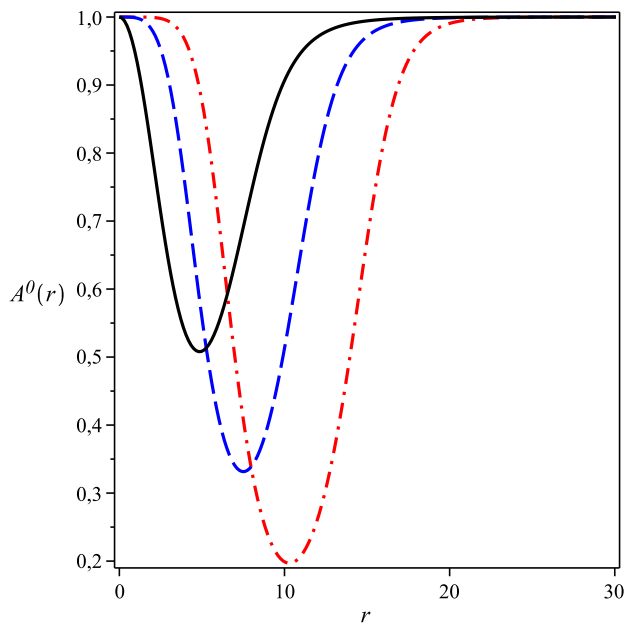}
\par
\vspace{-0.3cm}
\caption{Numerical solutions to $A^{0}(r)$. Conventions as in the Fig. 7.
This field behaves in the same way as before.}
\end{figure}

It is instructive to highlight that, in view of the analytical results we
have obtained, the energy-bound (\ref{eb}) can be calculated explicitly, its
approximate value being given by%
\begin{equation}
E_{bps}=\mp 8\pi h\left( m+1\right) \text{,}
\end{equation}%
the magnetic flux (\ref{xxmf2}) standing for%
\begin{equation}
\Phi _{B}=-\frac{8\pi }{g}(m+1)\text{,}
\end{equation}%
from which one gets $E_{bps}=\pm gh\Phi _{B}$, the energy of the analytical
first-order vortices being then proportional to their magnetic flux,
therefore verifying our previous conclusion established right after the Eq. (%
\ref{xxmf2}). We also point out that both the energy and the magnetic flux
of those vortices attained numerically are proportional to the effective
values of $A_{m}(r\rightarrow \infty )$.

\textit{The first-order solutions: the full numerical case}. It is important
to clarify that, beyond the configurations we have presented above, there is
a second first-order scenario which can not be predicted by any analytical
construction, i.e. it is not possible to approximate its solutions via $%
\alpha \left( r\right) \ll 1$. In order to introduce these new solutions, we
again solve the first-order equations (\ref{xbps11}) and (\ref{xbps22})
numerically according the conditions (\ref{abcx0}) and (\ref{abcx}), from
which we depict the resulting profiles in the figures 7, 8, 9, 10, 11 and 12
below. Here, we use $h=g=\kappa =r_{0}=1$ and $m=1$ (solid black line), $m=2$
(dashed blue line) and $m=3$ (dash-dotted red line).

The new numerical solutions for $\alpha \left( r\right) $ are depicted in
the Figure 7, the resulting configurations behaving in the same way before,
i.e. being rings centered at the origin whose radii and amplitudes
increasing as the vorticity $m$ itself increases.

In the Figure 8, we show the profiles to the gauge function $A\left(
r\right) $, the additional dotted orange line representing the solution for $%
m=4$. Here, it is important to point out the existence of an interesting
internal structure inherent to the new gauge profiles. Moreover, we
emphasize that the new solutions do not obey $A\left( r\rightarrow \infty
\right) \rightarrow 4(m+1)$, as expected.
\begin{figure}[tbp]
\centering\includegraphics[width=8.5cm]{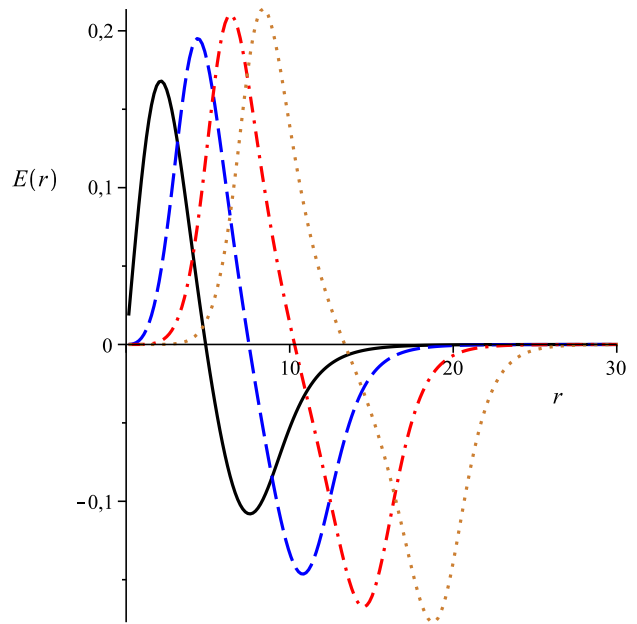}
\par
\vspace{-0.3cm}
\caption{Numerical solutions to $E(r)$. Conventions as in the Fig. 8. Note
the internal structure inherent to the electric field.}
\end{figure}

The solutions to the magnetic field $B(r)$ and the energy density $%
\varepsilon _{bps}\left( r\right) $ appear in the figures 9 and 10,
respectively, both ones standing for double rings centered at $r=0$. In
particular, the magnetic field satisfies $B(r=0)=0$ and $B(r\rightarrow
\infty )\rightarrow 0$, the energy distribution vanishing at the origin for $%
m\neq 1$ only, with $\varepsilon _{bps}(r\rightarrow \infty )\rightarrow 0$
(i.e. the finite-energy requirement still holds).

Finally, the figures 11 and 12 show the numerical solutions to the electric
potential $A^{0}(r)$ and the electric field $E(r)$, from which we see that
these two fields behave in the same way as those depicted in the figures 5
and 6 (including the sign inversion inherent to the electric field),
respectively, the electric one also possessing an internal structure, see
the dotted orange line.

%%%%%%%%%%%%%%%%%%%%%%%%

\subsection{The BPS formalism for $\protect\beta (r)=\protect\beta _{2}
\displaystyle$}

We now summarize the implementation of the BPS formalism for the case%
\begin{equation}
\beta \left( r\right) =\beta _{2}=\frac{\pi }{2}k\text{,}  \label{bxxb}
\end{equation}%
which gives $\cos ^{2}(2\beta _{2})=1$\ and $W(\alpha ,\beta _{2})=%
\displaystyle\frac{1}{4}\sin ^{2}2\alpha $.

In this case, the total energy obtained from (\ref{edx}) reads%
\begin{eqnarray}
E &=&2\pi h\int_{0}^{\infty }\left[ \left( \frac{d\alpha }{dr}\right) ^{2}+%
\frac{\sin ^{2}2\alpha }{4r^{2}}\left( \frac{A}{2}-m\right) ^{2}\right] rdr
\notag \\
&&+2\pi \int_{0}^{\infty }\left[ \frac{4\kappa ^{2}B^{2}}{g^{2}h\sin
^{2}2\alpha }+V\right] rdr\text{.}
\end{eqnarray}

Moreover, after some algebraic manipulation similar to the one we have
performed in the case $\beta \left( r\right) =\beta _{1}$, we attain the
following condition for the self-interacting potential%
\begin{equation}
\frac{4\kappa }{g^{2}\sqrt{h}}\frac{d}{d\alpha }\left( \frac{\sqrt{V}}{\sin
\left( 2\alpha \right) }\right) =\frac{h}{4}\frac{d}{d\alpha }\cos \left(
2\alpha \right) \text{,}
\end{equation}%
which can be solved to give%
\begin{equation}
V\left( \alpha \right) =\frac{g^{4}h^{3}}{1024\kappa ^{2}}\sin ^{2}\left(
4\alpha \right) \text{.}  \label{sdx1}
\end{equation}

\textbf{I}n view of this result, the total energy (\ref{cdx2}) can be
written as
\begin{eqnarray}
E &=&E_{bps}+2\pi h\int_{0}^{\infty }\left[ \frac{d\alpha }{dr}\mp \frac{%
\sin \left( 2\alpha \right) }{2r}\left( \frac{A}{2}-m\right) \right] ^{2}rdr
\notag \\
&&+2\pi \int_{0}^{\infty }\left( \frac{2\kappa B}{g\sqrt{h}\sin \left(
2\alpha \right) }\mp \frac{g^{2}h^{3/2}}{32\kappa }\sin \left( 4\alpha
\right) \right) ^{2}rdr\text{,}
\end{eqnarray}%
where the lower-bound now reads%
\begin{equation}
E_{bps}=2\pi \int r\varepsilon _{bps}dr=\mp \pi \frac{h}{2}A_{\infty }\text{,%
}
\end{equation}%
which is saturated when the profile fields satisfy%
\begin{equation}
\frac{d\alpha }{dr}=\pm \frac{\sin \left( 2\alpha \right) }{2r}\left( \frac{A%
}{2}-m\right) \text{,}
\end{equation}%
\begin{equation}
B=\pm \frac{g^{3}h^{2}}{64\kappa ^{2}}\sin \left( 2\alpha \right) \sin
\left( 4\alpha \right) \text{.}
\end{equation}

We point out that also the potential in (\ref{sdx1})\ can be written as an
explicit function of $\left\vert \phi _{3}\right\vert $, i.e.%
\begin{equation}
V\left( \left\vert \phi _{3}\right\vert \right) =\frac{g^{4}}{64k^{2}h}%
\left\vert \phi _{3}\right\vert ^{2}\left( h-\left\vert \phi _{3}\right\vert
^{2}\right) \left( h-2\left\vert \phi _{3}\right\vert ^{2}\right) ^{2}\text{,%
}
\end{equation}%
which manifestly breaks the original $SU(3)$\ symmetry, as expected.

Here, it is important to highlight that a simple comparison reveals that the
first-order results obtained for $\beta \left( r\right) =\beta _{2}$\ can be
mapped directly from those calculated for $\beta \left( r\right) =\beta _{1}$%
\ via the redefinitions $\alpha \rightarrow 2\alpha $ and $h\rightarrow h/4$.

\vspace{.5cm}

%%%%%%%%%%%%%%%%%%%%%%%%

\section{Final comments}

\label{Intro copy(1)}

In this work, we have considered the nontopological first-order solitons
inherent to a planar gauged $CP(2)$ scenario endowed by the Chern-Simons
action, focusing our attention on those time-independent profiles possessing
radial symmetry. We have proceeded the minimization of the corresponding
energy (the starting-point being the energy-momentum tensor), from which we
have established the corresponding first-order framework (a set of two
coupled first-order equations and a well-defined lower bound for the total
energy itself) inherent to the effective radially symmetric scenario.

In the sequel, we have solved the first-order equations numerically by means
of a finite-difference method. In this sense, despite the high nonlinearity,
we have identified a special kind of configurations that can be described by
approximate analytical solutions in the regimen $\alpha (r)\ll 1$ for all $r$%
.\ The resulting profiles have been depicted and we have commented their
main characteristics, from which we have noted an interesting inversion of
the sign dictating the electric interaction and the existence of an internal
structure inherent to the gauge function.

An interesting issue for a future work includes the search for a more
general implementation of the first-order BPS formalism independent of an
specific Ansatz. This idea is currently being under investigation and we
hope positive results to be presented in an incoming contribution.

\begin{acknowledgments}
This work was supported by the CNPq, CAPES and FAPEMA (Brazilian agencies).
In particular, RC thanks the support from the grants CNPq/306385/2015-5, FAPEMA/Universal-00782/15 and FAPEMA/Universal-01131/17, MLD acknowledges the full support from CAPES (postgraduate scholarship), and EH thanks the support from the grants CNPq/307545/2016-4 and CNPq/449855/2014-7.
\end{acknowledgments}

\end{document}